\documentclass[twocolumn,prb,showpacs]{revtex4}
\usepackage{graphicx}
\usepackage[pdfpagemode=FullScreen,bookmarks=true]{hyperref}

\begin{document}

\title{Ising Dynamics with Damping}

\author{J. M. Deutsch}
\affiliation{ Department of Physics, University of California, Santa Cruz, CA 95064.}
\author{A. Berger}
\affiliation{CIC Nanogune, Mikeletegi Pasealekua 56, 301 E-20009 Donostia Spain}

\begin{abstract}
We show for the Ising model that is possible  construct a discrete time
stochastic model analogous to the Langevin equation that incorporates an
arbitrary amount of damping.  It is shown to give the correct equilibrium
statistics and is then used to investigate nonequilibrium phenomena, in
particular, magnetic avalanches. The value of damping can greatly alter
the shape of hysteresis loops, and for small damping and high disorder,
the morphology of large avalanches can be drastically effected. Small
damping also alters the size distribution of avalanches at criticality.
\end{abstract}


\maketitle

\section{Introduction}

In many situations, it is useful to discretize continuous degrees of
freedom to better understand them, both from a theoretical standpoint and
for numerical efficiency.  Ising models are perhaps the best example of
this and have been the subject of numerous theoretical and numerical
studies. Renormalization group arguments~\cite{SKMaCritPhenom}  have explained the reason
why this discretization  gives equilibrium critical properties of many
experimental systems, and these kinds of arguments have been extended to
understanding their equilibrium dynamics~\cite{HalperinHohenbergMa1}.  For non-equilibrium situations,
such as the study of avalanches, such arguments probably do also apply
to large enough length and time scales as well. However there are many
situations where it would be desirable to understand smaller length
scales where other factors should become relevant.

This is particularly true with dynamics of magnetic systems, where
damping is often weak in comparison to precessional effects. For
studies of smaller scales, it has been necessary to use more time
consuming micromagnetic simulations utilizing continuous degrees of
freedom, such as the Landau Lifshitz Gilbert equations~\cite{LLGref}
which is a kind of Langevin equation that gives the stochastic evolution
of Heisenberg spins. 
\begin{equation}
\frac{d{\bf s}}{dt} = -{\bf s} \times ({\bf B} 
-\gamma {\bf s} \times {\bf B}),
\label{eq:LLG}
\end{equation}
where ${\bf s}$ is a microscopic magnetic moment, ${\bf B}$ is the
local effective field, and $\gamma$ is a damping factor,
measuring the relative importance of damping to precession.  In real
materials it ranges~\cite{GammaValues} from small damping $\gamma = .01$,
to $1$. In contrast, the dynamical rules implemented for Ising models are
most often ``relaxational" so that energy is instantaneously dissipated
when a spin flips, as with the Metropolis algorithm.

However there is a class of ``microcanonical"  Ising dynamics~\cite{Creutz}
reviewed in section \ref{sec:NonRelaxDyn} where auxiliary degrees of freedom are
introduced and all moves conserve the total energy.  The other degrees of
freedom can be taken to be variables associated with each spin, and allowed
moves can change both the state of the spins and the auxiliary variables.  This
can be thought of crudely, as a discretized analogy to molecular dynamics, and
is also similar to discrete lattice gas models of
fluids~\cite{LatticeGasFrisch,LatticeGasZanetti}. These models give the correct
equilibrium Ising statistics of large systems and can also be used to understand
dynamics in a different limit than the relaxational case.

Real spin systems are intermediate between these two kinds of dynamics
and as mentioned above, are better described by Langevin dynamics. In
the context of spins, the question posed and answered here is: how
does one formulate a discrete time version of stochastic dynamics that
includes damping and gives the correct equilibrium statistics?  In section \ref{subsec:ExtenToDamping}
we are able to show that there is a fairly simple method for doing this using
a combination of microcanonical dynamics, and an elegant procedure that
incorporates damping and thermal noise.  This procedure differs from that
of the Langevin equation in that it requires non-Gaussian noise. Despite
this, the noise has surprisingly simple but unusual statistics.

We will then show that this procedure gives the correct equilibrium
statistics and verify this numerical in section \ref{subsec:EqTests}
with a simulation of the two dimensional Ising model with different
amounts of damping.

Because the value of damping is an important physical parameter in
many situations it is important that there is a straightforward way of
incorporating its effects in Ising simulations. This is particularly
noteworthy as Ising kinetics are a frequently used means of understanding
dynamics in many condensed matter systems.

After this in section \ref{subsec:AvalDyn}  we will turn to nonequilibrium problems where, using
this approach, we can study the effects of damping on a number of
interesting properties of systems displaying avalanches and Barkhausen
noise~\cite{Barkhausen}.  We first show how to modify the kinetics for
this case and then study systems in two and three dimensions. With modest
amounts of computer time, we can analyze problems that are out of the
reach of micromagnetic simulations and allow us to probe the effects
of damping on the properties of avalanches. This is related to recent
work~\cite{DeutschBerger1} by the present authors using both the Landau
Lifshitz Gilbert equation, Eqn. \ref{eq:LLG}, and theoretical approaches, to understand how
relaxational dynamics of avalanches~\cite{SethnaReview}, are modified at
small to intermediate scales by this more realistic approach. With the
present approach we find new features and modifications of avalanche
dynamics. We find that the shape of hysteresis loops can be strongly
influenced by the amount of damping. One of the most striking findings
is that there exists a parameter regime of high disorder and small
damping where single system-size avalanches occur that are made up of a
large number of disconnected pieces.  We can also analyze the critical
properties of avalanches when damping is small and give evidence that
there is a crossover length scale, below which avalanches have different
critical properties.

\section{Non-relaxational dynamics}
\label{sec:NonRelaxDyn}

We start by considering a model for a magnet with continuous degrees of freedom,
such as a Heisenberg model with anisotropy. The Ising approximation simplifies
the state of each spin to either up or down, that is $s_i = \pm 1$,
$i=1,\dots,N$. One important effect that is ignored by this approximation is
that of spin waves that allow the transfer of energy between neighbors, and for
small oscillations, give an energy contribution per spin equal to the
temperature $T$ (here we set $k_B =1$).  This motivates the idea that there are
extra degrees of freedom associated with every spin that can carry (a positive)
energy $e_i$.  Creutz introduced such degrees of freedom~\cite{Creutz} and
posited that they could take any number of discrete values. He used these
auxiliary variables $e_i$ to construct a cellular automota to give the correct
equilibrium statistics for the Ising model, in a very efficient way that did not
require the generation of random numbers.
Thus we have a Hamiltonian $H_{tot}$ that is the sum of both
spin $H_{spin}$ and auxiliary degrees of freedom 
$H_e$: $H_{tot} = H_{spin}+H_e$. $H_{spin}$ can be a general Ising spin Hamiltonian and 
$H_e = \sum_i e_i$. In our model there is a single auxiliary variable $e_i$ associated with each
lattice site $i$, that can take on any real value $\ge 0$.

However for the purposes of trying to model dynamics of spins, it also makes
sense to allow the $e_i$'s to interact and exchange energy between neighbors.
For example, one precessing spin should excite motion in its neighbors. This
exchange was formulated in the context of solidification using a Potts model
instead of an Ising model by Conti {\em et al.}~\cite{Conti}, but can equally
well be used here. 

Now we
can formulate a microcanonical algorithm for the Ising model using a
procedure very similar to their prescription. In each step:

\begin{itemize}
\item [1.] We choose a site $i$ at random.
\item [2.] We randomly pick with equal probability either a spin or an
auxiliary degree of freedom, $s_i$ of $e_i$:
\begin{itemize}
\item [(a)] $s_i$'s: We attempt to move spins (such as the flipping
      of a single spin). If the energy cost in doing this is $\leq e_i$
      we perform the move and decrease $e_i$ accordingly. Otherwise
      we reject the move.
\item [(b)] $e_i$'s: We pick a nearest neighbor $j$, and repartition 
              the total energy with uniform probability between these two variables.
              That is, after repartitioning, $e_i' = (e_i+e_j) r$ and 
$e_j' = (e_i+e_j) (1-r)$, where $0 < r < 1$ is uniform random variable.
\end{itemize}
\end{itemize}
Note that these rules preserve
the total energy and the transitions between any two states have the
same probability. Therefore this will give the correct microcanonical
distribution. For large $N$, this is, for most purposes~\cite{Lax},
equivalent to the canonical distribution $\propto \exp{(-\beta
H_{tot})}$. Note that the probability distribution for each variable
$e_i$, $P(e_i) = \beta \exp{(-\beta e_i)}$, so that the $\langle
e_i\rangle = T$.  That is, measurement of average of $e_i$'s directly
gives the effective temperature of the system.

\subsection{Extension To Damping}
\label{subsec:ExtenToDamping}

The question we asked, is how to extend this equilibrium simulation method
to include damping.  In this case the system is no longer closed and
energy is exchanged with an outside heat bath through interaction with
the auxiliary variables.  As with the Langevin equation, there are two
effects. The first is that the energy is damped. Call the dissipation
parameter for each step $\alpha$, which will lie between $0$ and $1$.
Then at each time step we lower the energy with $e_i \rightarrow \alpha
e_i$ for all sites $i$. By itself, this clearly will not give a system
at finite temperature and we must also include the second effect of a
heat bath, which adds energy randomly to the system.  In the case of
the Langevin equation, a Gaussian noise term $n(t)$ is added to keep
the system at finite temperature. A discretized version of this, that
evolves the energy $e(t)$ at time step $t$ is
\begin{equation}
e(t+1) = \alpha e(t) + n(t).
\end{equation}
This equation will not work if the noise  $n(t)$ is Gaussian as this does
not give the Gibbs distribution $P_{eq}(e) = \beta \exp{(-\beta e)}$.
Therefore we need to modify the statistics of $n(t)$. It is possible to
do so if we choose $n(t)$ at each time $t$ from a distribution
\begin{equation}
p(n) = \alpha \delta(n) + (1-\alpha) \beta e^{-\beta n} \theta(n)
\label{eq:p(n)}
\end{equation}
where $\theta$ is the Heaviside step function.  To show this, we write
down the corresponding equation for the evolution of the probability
distribution for $e$:

\begin{widetext}
\begin{equation}
   \label{eq:EvolOfP}
P(e',t+1) = \langle \delta (e'-(\alpha e +n))\rangle = \int \int P(e,t) p(n) \delta (e'-(\alpha e +n)) de~dn
\end{equation}
\end{widetext}
We require that the $P$ as $t \rightarrow \infty$ obeys $P(e,t+1) = P(e,t) =
P_{eq}(e) = \beta \exp(-\beta e)$, for $e > 0$.  It is easily verified
that by choosing this form of $P(e,t)$ and
by choosing $P(n)$ as in Eq. \ref{eq:p(n)}, we satisfy Eq. \ref{eq:EvolOfP}.

Therefore to add damping to this model, we add the following procedure
to the steps stated above:
\begin{itemize}
\item [3.]
Choose a uniform random number $0 < r < 1$. If $r < \alpha$, then $e_i
\rightarrow \alpha e_i$.  Otherwise $e_i \rightarrow \alpha e_i - T \ln (r')$, 
where $r'$ is another uniform random number between $0$ and $1$.
\end{itemize}

If we assume that the probability distribution for the total system is of the form 
$P_{Gibbs} \propto \exp{(-\beta H_{tot})} = \exp{(-\beta H_{spin})}\exp{(-\beta H_e)}$, 
we will now show that the steps 1, 2, and 3, of this algorithm preserve this distribution.
Following the same reasoning as above for the microcanonical simulation,
moves implementing steps 1 and 2 do not change the total energy, and they
preserve the form of $P_{Gibbs}$ because $P_{Gibbs}$ depends only on the
total energy ($H_{tot}$), and 1 and 2 explore each state in an energy shell with
uniform probability. Because of the form of $P_{Gibbs}$, its dependence
on the variable $e_i$ is $\propto \exp{(-\beta e_i)}$. According to the
above argument, after step 3, it will remain unchanged. Therefore all
steps in this algorithm leave $P_{Gibbs}$ unchanged. The algorithm
is also ergodic, and therefore this will converge to the Gibbs
distribution~\cite{SethnaBook} as $t \rightarrow \infty$.

Because the steps each  preserve the Gibbs distribution, the ordering of
them is not important in preserving equilibrium statistics. For example,
we could sweep through the lattice sequentially instead of picking $i$
at random. We could perform step 3 after steps 1 and 2 were performed
$N$ times.

\subsection{Equilibrium Tests}
\label{subsec:EqTests}

\begin{figure}[htp]
\includegraphics[width=\hsize]{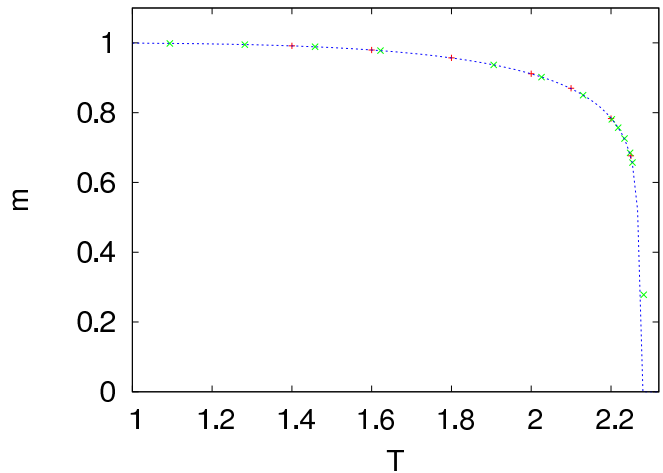}
\caption{
Plot of results obtained for the two dimensional Ising model on a
$128^2$ lattice for
two different values of the damping parameter. This is a plot of the
average magnetization per spin $m$ vs. $T$. The $\times$'s are
for no dissipation, $\alpha =1$, which is a purely microcanonical
simulation. The $+$'s are for $\alpha = 0.5$. The dashed curve is
the exact solution to this model in the thermodynamic limit.
}
\label{fig:m_vs_T}
\end{figure}

\begin{figure}[htp]
\includegraphics[width=\hsize]{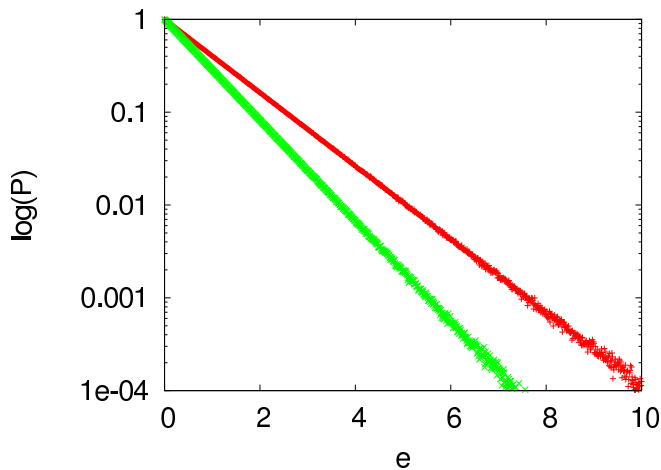}
\caption{
Plot of results obtained for the two dimensional Ising model on a
$128^2$ lattice for the probability distribution for the auxiliary variables
$e_i$, at two different temperatures with a damping parameter $\alpha =0.5$.
The upper curve is for $T=1.1$ and the lower for $T=0.8$.
}
\label{fig:histe}
\end{figure}

We performed tests on this algorithm and verified that it did indeed
work as expected.  We simulated the two dimensional Ising model on
a $128^2$ lattice with different values of the damping parameter,
and compared it with the exact results.  The average magnetization
per spin $m$ is plotted in Fig.~\ref{fig:m_vs_T} as a function of
the temperature $T$ and compared with the exact result~\cite{McCoyWu} 
for large $N$ (dashed curve).  The $\times$'s are the case $\alpha = 1$, which is
then just an implementation of the microcanonical method~\cite{Conti}
described above. In this case, the temperature was obtained by measuring
$\langle e_i\rangle$ because the energy was fixed at the start of the
simulation. The only point which is slightly off the exact solution is
in the critical region, as is to be expected.  The case $\alpha = 0.5$
is shown with the $+$'s and lie on the same curve. Results were obtained
for $\alpha = 0.9$ but are so close as to be indistinguishable and are
therefore not shown.  We also checked that the distribution of auxiliary
variables had the correct form. The probability distribution for the
energy $e$ is shown in Fig. \ref{fig:histe}.  Fig. \ref{fig:histe} plots
the distribution $P(e_i)$ versus energy $e_i$, averaged over all sites
$i$ on a linear-log scale for $\alpha = 0.5$ and $T = 0.8$,  and $1.1$.
The curves are straight lines over four decades and show the correct
slopes, for $T = 0.8$, $\langle e_i \rangle = 0.8002$ and for $T = 1.1$,
$\langle e_i \rangle = 1.1003$.

\section{Avalanche dynamics}
\label{subsec:AvalDyn} 

Avalanche dynamics of spin systems have been mainly studied using models
that are purely relaxational. There is a whole range of interesting
phenomena that have been elucidated by such studies and have yielded
very interesting properties. The simplest model that can be used in
this context is the random field Ising model (RFIM) with a Hamiltonian
\begin{equation}
{\cal H} = -\sum_{<ij>} J s_i s_j  - \sum_{i} h_i s_i - h \sum s_i
\end{equation}
where $J$ is the strength of the nearest neighbor coupling, $h_i$
is a random field, with zero mean, and $h$ is an externally applied
field.  A magnet is placed in a high field $h$ and then this is
very slowly lowered. As this happens, the spins will adjust to the
new field by flipping to lower their energy. In the usual situation,
the system is taken to be at $T=0$, so that only moves that lower the
energy are accepted.  The flipping of one spin can cause a cascade of
additional spins to flip, causing the total magnetization $M$ to further
decrease. The occurrence of these cascades is called an ``avalanche".
At zero temperature there is one parameter $j$ that characterizes the
system, the ratio of nearest neighbor coupling to the distribution width of the
random field. One considers the behavior of a system when its starts in
a high field and is slowly lowered.  When $j$ is small the system is
strongly pinned and the system will have a number of small avalanches
generating a smooth hysteresis loop. For large $j$, the system will
have a system-size avalanche involving most of the spins in the system,
leading to a precipitous drop in the hysteresis loop. There is a critical
value of $j$ where the distribution of avalanche sizes is a power law
and self-similar scaling behavior is observed.

Here we investigate how this is modified by adding damping to
these zero temperature dynamics according to the following
rules: 
\begin{itemize}
\item [1.] The field is slowly lowered by finding the next field where 
a spin can flip.
\item[2.] The spins then flip, exchanging energy with auxiliary variables
$e_i$ as described above. The number of times this is attempted is $n_m$ times 
the total number of spins in the system. Here we set $n_m =16$. In more detail:
\begin{itemize}
\item [(i)] {\bf Spin moves:} An attempt to move each spin on the lattice 
is performed by attempting to  flip sequentially every third spin, in order 
to minimize artifacts in the dynamics due to updating contiguous spins. 
(The lattice sites are linearly ordered using ``skew" boundary conditions). 
Then all three sublattices are cycled over.
\item[(ii)]  {\bf Energy moves:} Exchange of energy with nearest neighbors is
performed cycling through all directions of nearest neighbors. 
Using the same sequence of updates, the $e_i$'s  exchange
energy with their nearest neighbors in one particular direction.
\item[(iii)]  {\bf Dissipation:} The energy of each $e_i$ is lowered to
   $\alpha e_i$.
\end{itemize}
\item[3.] We check for when the spins have settled down as follows:
if the $e_i$'s are not all below some energy threshold $e_{thresh}$,
set below to be $10^{-4}$, or the spin configuration has changed, step
2 is repeated until these conditions are both met. 
\item[4.] When the spins have settled down, we go to step 1.
\end{itemize}
The parameters $n_m$ and $e_{thresh}$ were varied to check that the correct
dynamics were obtained. The larger $\alpha$, the smaller the dissipation
and the larger the number of iterations necessary to achieve the final static 
configuration.

\begin{figure}[htp]
\begin{center}
\includegraphics[width=\hsize]{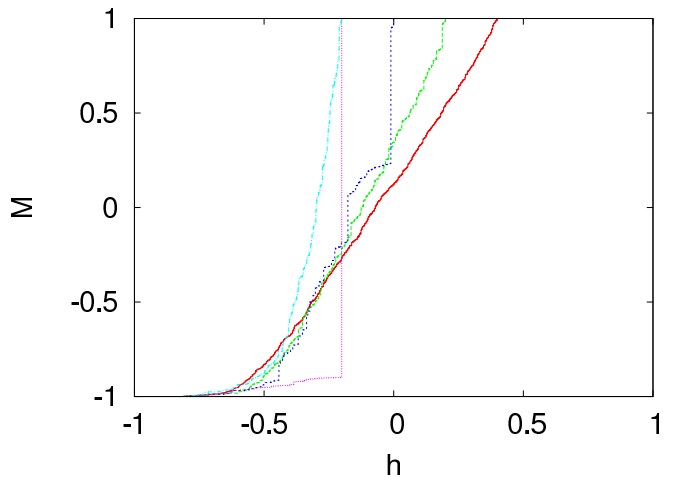}
\end{center}
\caption{
The major branch of the descending hysteresis loop for $64^2$ systems using
different values of the damping parameter and the spin coupling. Strong damping,
$\alpha = 0.5$ is shown in the left most curve (as judged from the top of the
plot) for coupling $j=0.3$ which starts decreasing from $M=1$ at $h= -0.2$, and
does not have large abrupt changes.  All the other curves are for weak damping,
$\alpha = 0.99$.  In this case but also for $j = 0.3$, we see that although $M$
starts to decrease at the same location as for strong damping, it drops abruptly
as the field is lowered.  As the coupling $j$ is decreased, smooth curves are
eventually seen again.  Going left to right, as judged from the top, are
$j=0.3$, $0.25$, $0.2$, and $0.15$.
}
\label{fig:m_vs_h_99_2d}
\end{figure}

\begin{figure*}[htp]
\begin{center}
(a)
\includegraphics[width=0.4 \hsize]{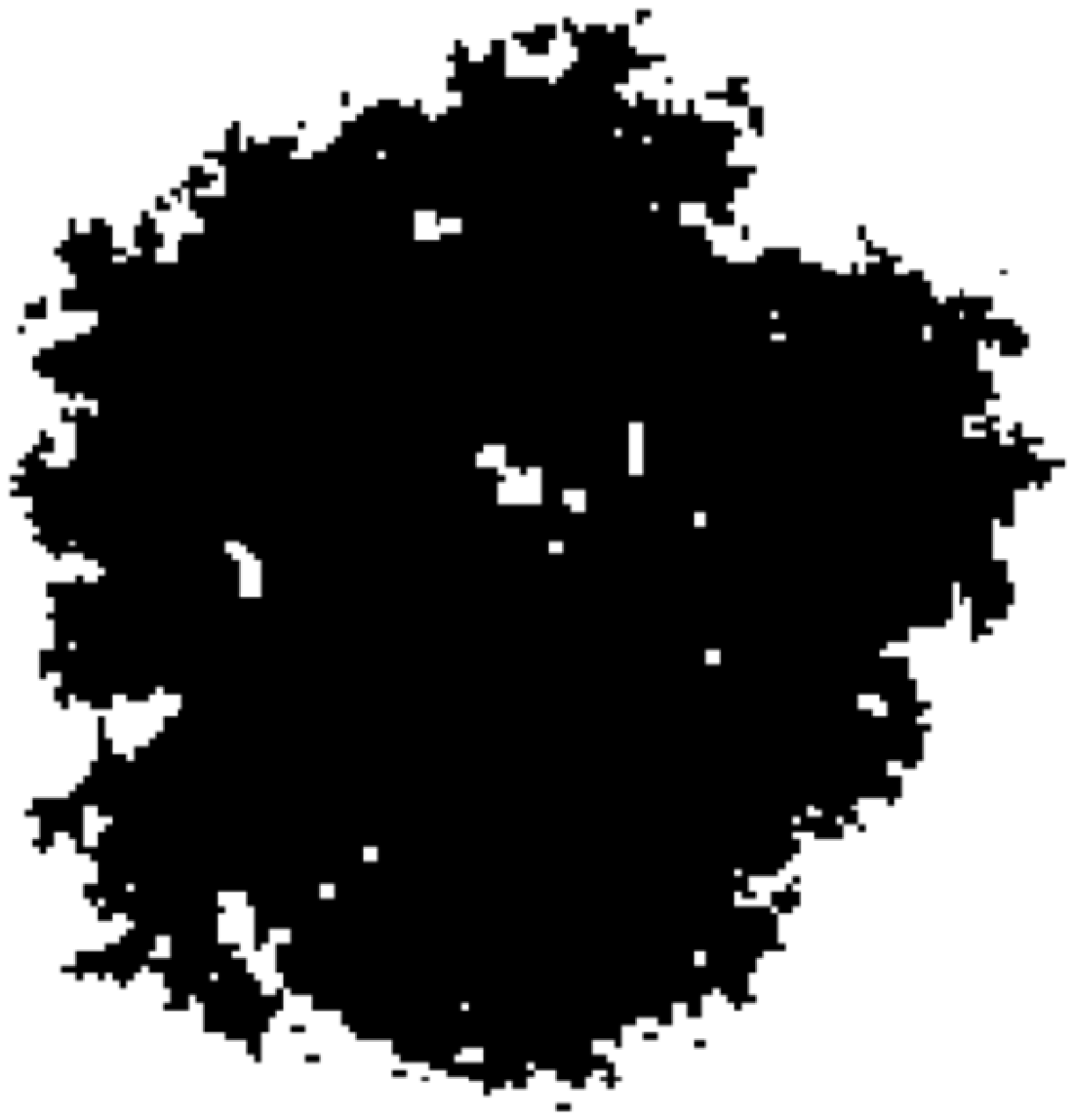}
(b)
\includegraphics[width=0.4 \hsize]{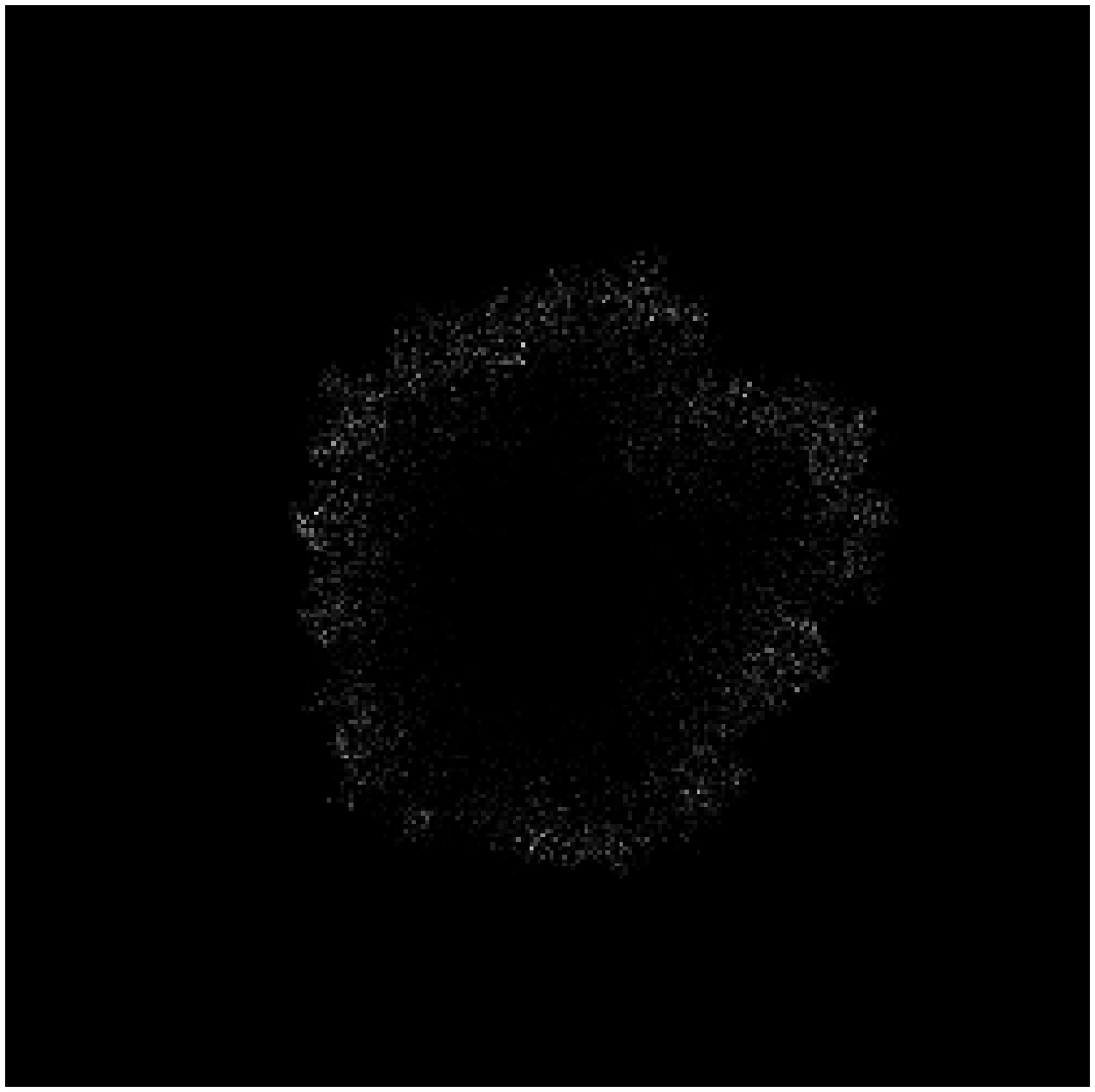}
\end{center}
\caption{
(a) The spin configuration for a $256^2$ system with $j = 0.35$,
$\alpha = 0.9$ during a system size avalanche at the field $h= -0.400007$. (b) A gray-scale
plot of the auxiliary variables at the same time.
}
\label{fig:config_n35_diss9}
\end{figure*}

\begin{figure*}[htp]
\begin{center}
(a)
\includegraphics[width=0.3 \hsize]{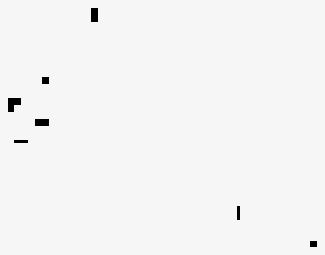}
(b)
\includegraphics[width=0.3 \hsize]{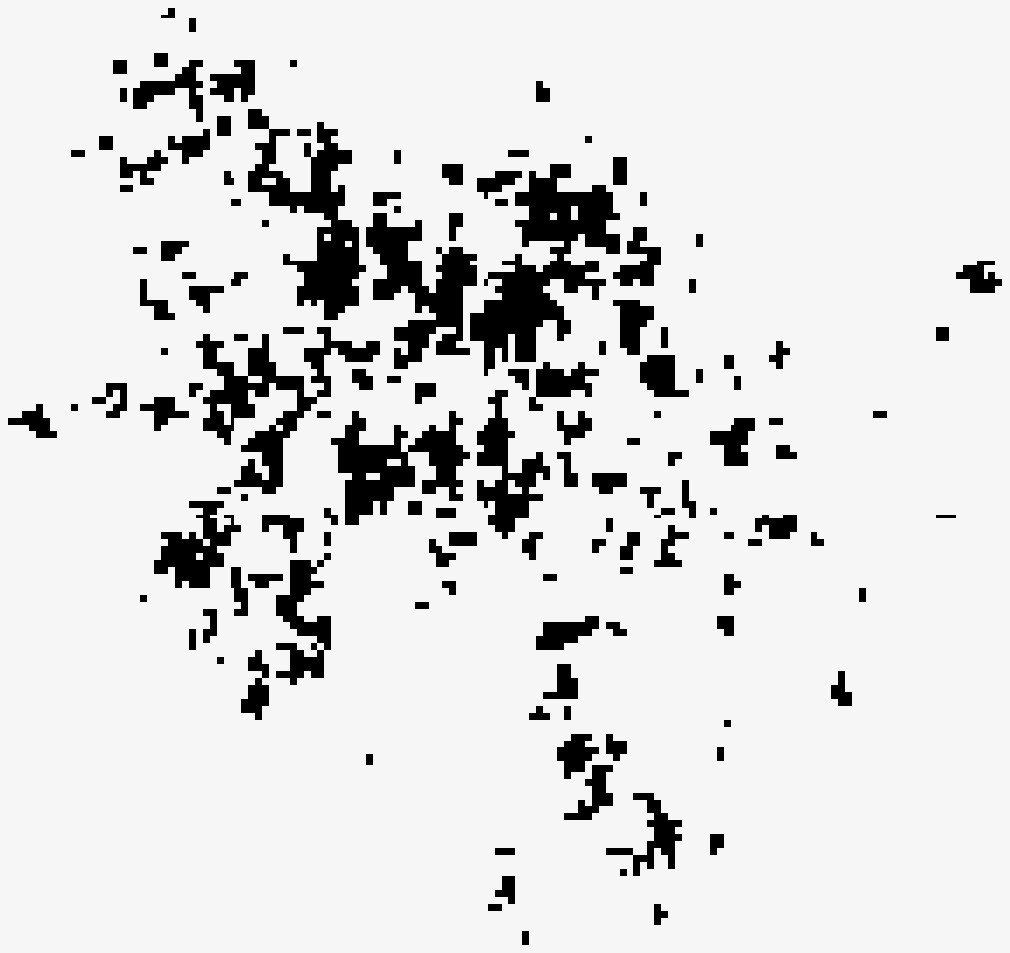}
(c)
\includegraphics[width=0.3 \hsize]{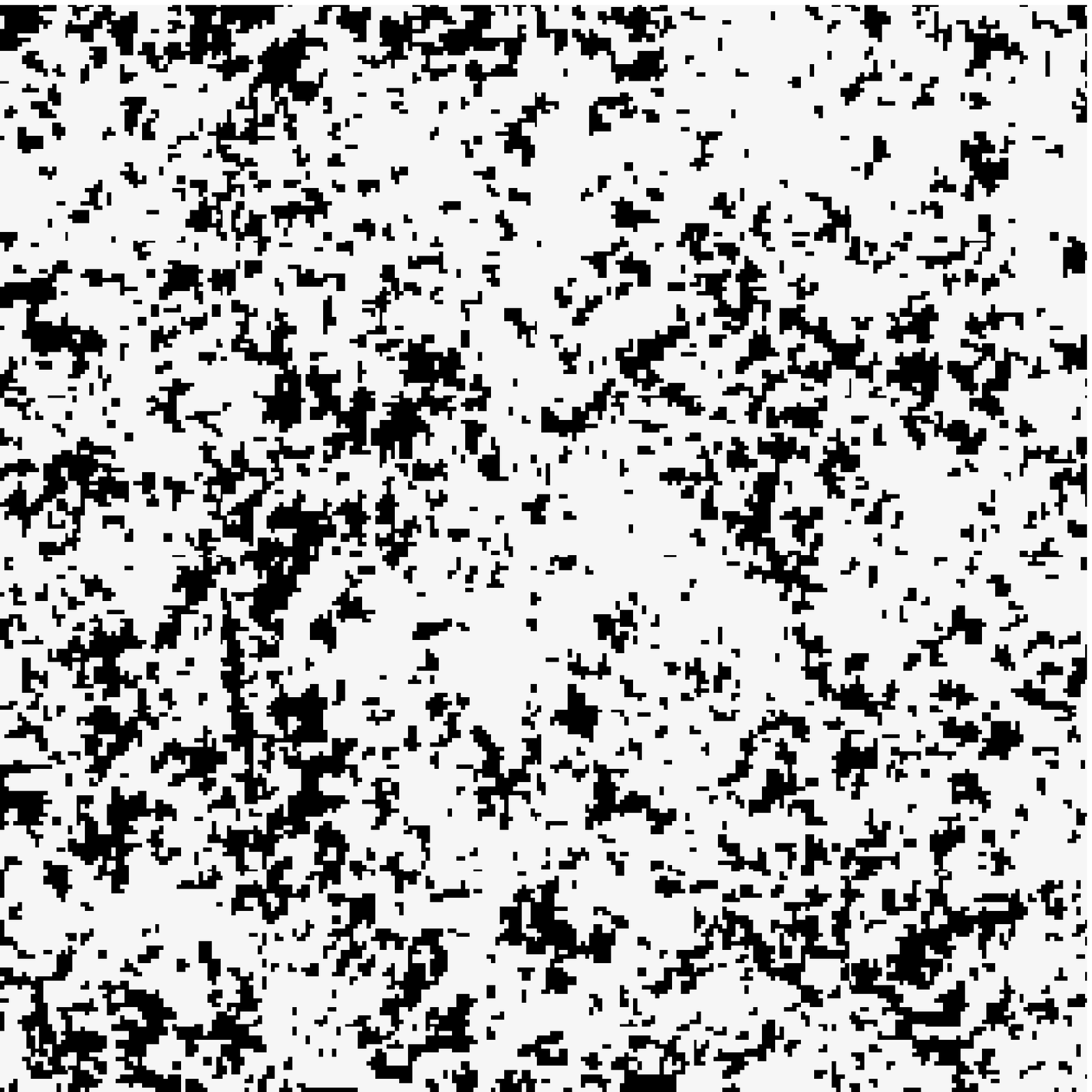}
\end{center}
\caption{
Spin configurations for a $256^2$ system with $j = 0.25$, $\alpha =
0.99$ during an avalanche at the field $h= -7\times 10^{-5}$ . (a) The
beginning of the avalanche. (b) When the avalanche is of order of half
the system size.
(c) The final configuration of the avalanche.
}
\label{fig:config_n25_diss99}
\end{figure*}

\subsection{Two Dimensional Patterns}

We first investigate the case of two dimensions where it is simpler to
visualize the avalanches in various conditions than in three dimensions.
Much experimental work and theoretical work on avalanches has been
done on two dimensional magnetic films and this case should be highly
relevant~\cite{SethnaReview}.

We first examine how the hysteresis loops change as a function of the
coupling $j$ and the damping parameter $\alpha$ for a $64^2$ system.  The
major downwards hysteresis loops are shown in Fig. \ref{fig:m_vs_h_99_2d}
for a variety of parameters described below.  We first examine strong
damping $\alpha = 0.5$. For $j=0.3$ the hysteresis curve is quite smooth
with all avalanches much less than the system size (left most curve). Now
consider the same value of $j$ but with with small damping, $\alpha =
0.99$. The curve now is a single downwards step with a small tail at
negative $h$. The lower damping has allowed that system to form a system
size avalanche. The difference is due to the fact that with small damping,
the energy of avalanched spins is not immediately dissipated and as a
consequence, heats up neighboring spins, allowing them to more easily
avalanche as well.  Therefore a system size avalanche is seen in the small
damping case, leading to the precipitous drop in the hysteresis loop.

When the value of the coupling $j$ is lowered to $0.15$ for  $\alpha
= 0.99$, smooth loops are obtained.  The Fig. \ref{fig:m_vs_h_99_2d}
shows intermediate values of the coupling parameter as well.

To better understand the reason why the energy of the auxiliary variables
can trigger further spins to flip, in Fig. \ref{fig:config_n35_diss9} we
show the state of a system during a system size avalanche for $j = 0.35$
and a moderately small damping value, $\alpha= 0.9$, with $h= -0.400007$.
Fig. \ref{fig:config_n35_diss9}(a) shows that the flipped spins form
a fairly compact cluster and Fig. \ref{fig:config_n35_diss9}(b) shows
the corresponding values of the $e_i$'s in a gray scale plot, suitably
normalized. It has the appearance of a halo around the growth front
of the avalanche. The spins in the growth front have just flipped and
so energy there has not had a chance to diffuse or dissipate and so
has a higher spin temperature.  The interior is cold because damping
has removed energy from the auxiliary degrees of freedom. This higher
temperature diffuses into the the unflipped region allowing spins to
flip by thermal activation.

Because large avalanches are possible for small damping in a
parameter range where the relative effect of the random field
is much larger, it is of interest to see if avalanches have a
different morphology than typical large avalanches for high damping
systems. Fig. \ref{fig:config_n25_diss99} shows such spin configurations
first at the beginning of the avalanche and further along during
propagation when it has reached roughly half the system size, and
finally when it has reached its final configuration and the maximum
auxiliary variable value is $< 4\times 10^{-4}$.  The morphology of this
is very different than what is seen for large avalanches with stronger
coupling, for example Fig. \ref{fig:config_n35_diss9}.  At very small
fields, in this figure $h=-7 \times 10^{-5}$, surface tension precludes
the formation of minority domains, but because disorder is large, there
will be many small regions where the local field is much stronger and these
will want to form downward oriented (black) domains. There is a finite
activation barrier to forming these that can only be overcome at finite
temperature. However the majority of the spins still strongly disfavor
flipping. But because damping is small, heat has a chance to diffuse
through these regions into the favorable regions, allowing disconnected
regions to change orientation by thermal activation. Note that we have
checked numerically that small damping with strong coupling also leads
to compact configurations, so disorder is an essential ingredient in
this new morphology.

\begin{figure}[htp]
\begin{center}
\includegraphics[width=\hsize]{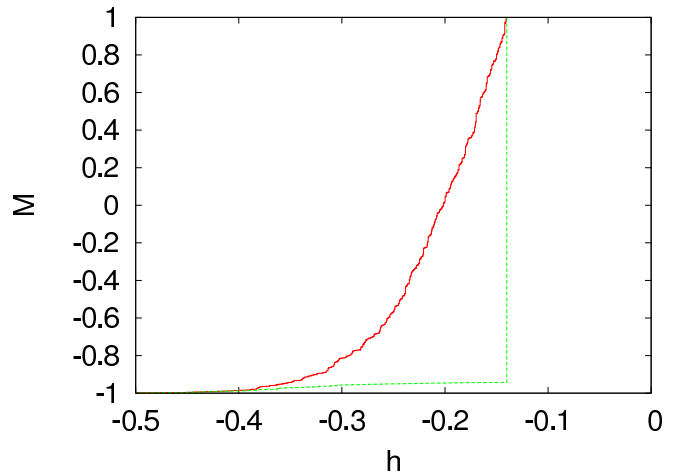}
\end{center}
\caption{
The major branch of the descending hysteresis loops
in two $32^3$ systems with $j=0.19$, for two different
values of the dissipation, upper curve: $\alpha = 0.5$,
lower curve: $\alpha = 0.99$.
}
\label{fig:comp_m_vs_h}
\end{figure}

\begin{figure*}[htp]
\begin{center}
(a)
\includegraphics[width=0.4 \hsize]{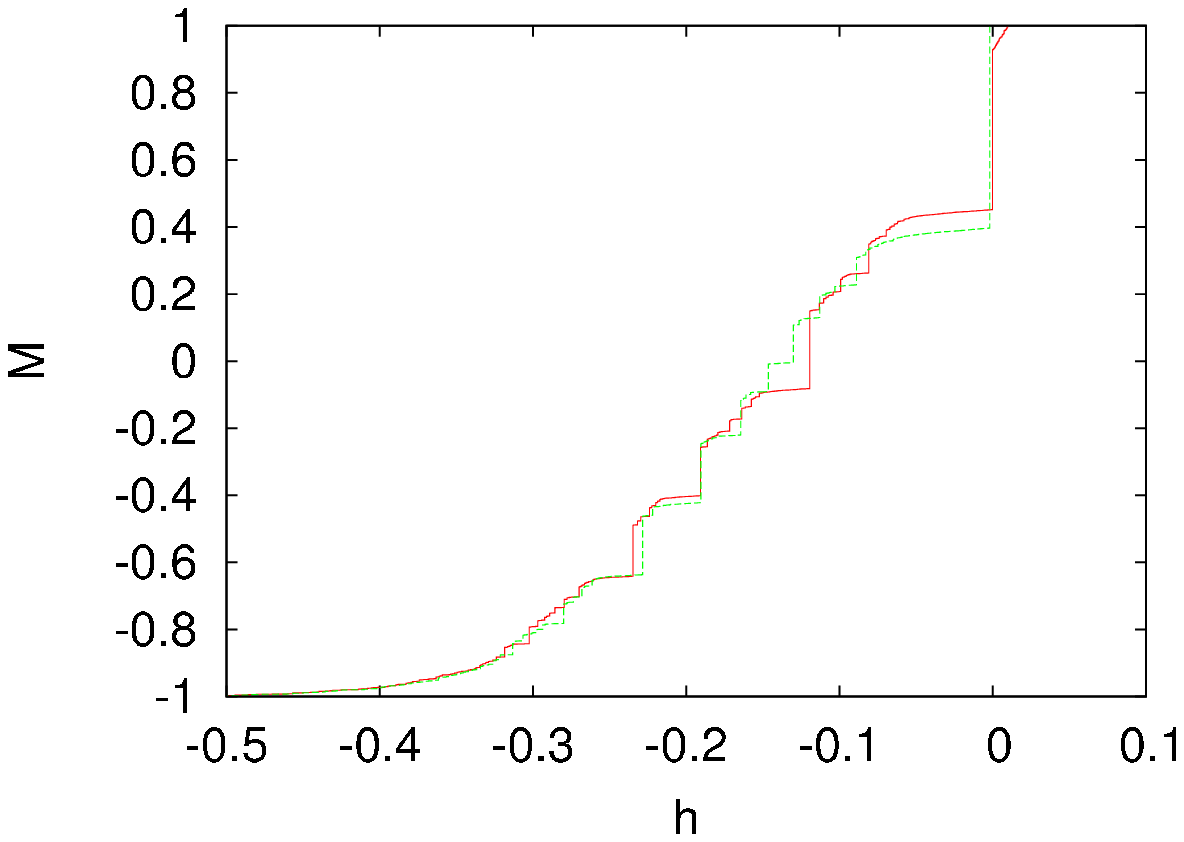}
(b)
\includegraphics[width=0.4 \hsize]{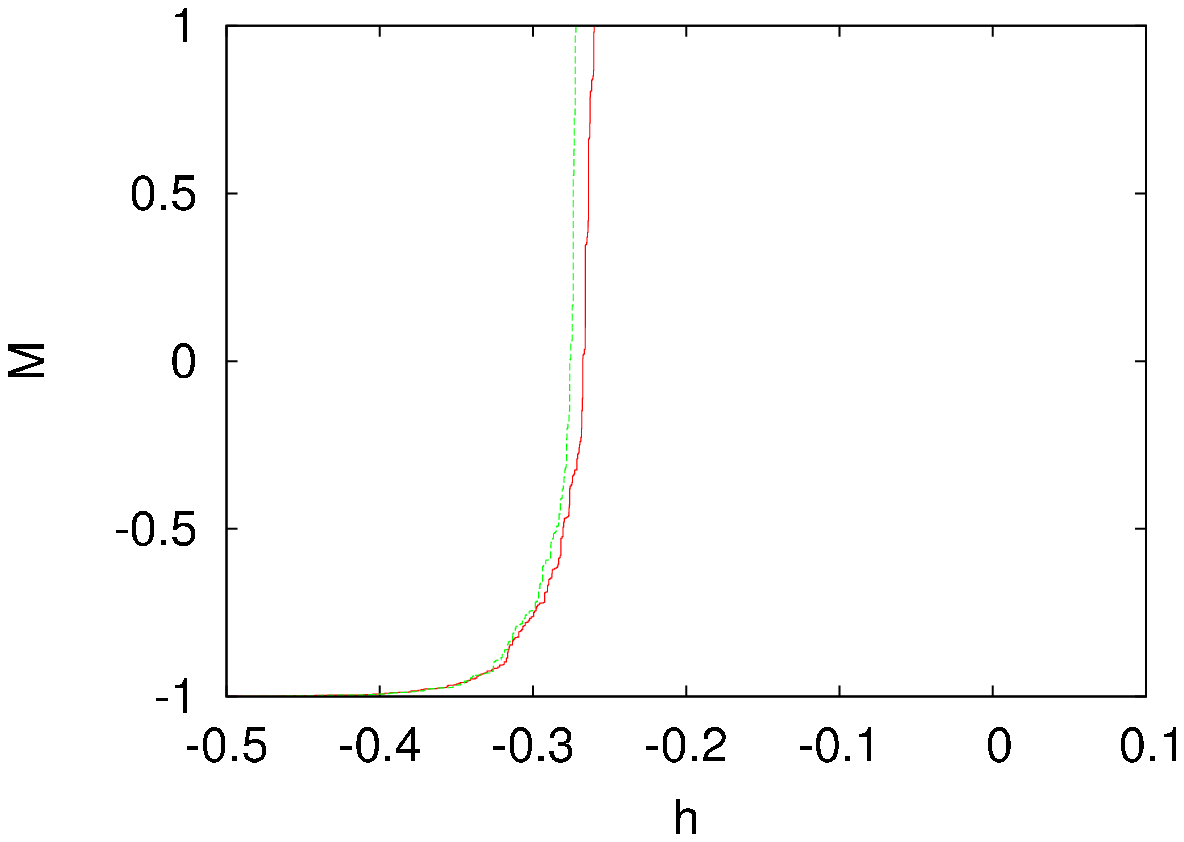}
\end{center}
\caption{
(a) Magnetization versus field for the Ising model with damping
described in the text. The system size is $32^3$ and the two lines
represent two runs close to criticality, one with a coupling of
$j=0.165$ and the other of $0.167$. (b) The plot for relaxational
dynamics (large damping) with couplings of $.21$ and $.212$. 
}
\label{fig:hyst_damping}
\end{figure*}

\subsection{Three Dimensions}

We first check that as with two dimensions, the value of the damping
parameter can have a large effect on the shape of a hysteresis loop.
Fig. \ref{fig:comp_m_vs_h} shows the downward branches of the major
hysteresis loop when the only parameter that is changed is the damping,
$\alpha$. The system is a $32^3$ lattice with $j = 0.19$. A value for
high damping, $\alpha = 0.5$, is the upper line. The lower line is for
small damping with $\alpha = 0.99$.

A more subtle effect, is that of damping on what happens near
criticality. In this case the value of the critical $j$ will
depend on the value of $\alpha$ as is apparent from the results of
Fig. \ref{fig:comp_m_vs_h}. At this point, the distribution of avalanche
sizes is expected to follow a power law distribution for large sizes. We
located this point and examined system properties in this vicinity.
Fig. \ref{fig:hyst_damping} shows
examples of such runs for $32^3$ systems.  Fig. \ref{fig:hyst_damping}(a)
shows a plot of the magnetization per spin $M$, versus the applied
field $h$ for $j = 0.165$ and $j= 0.167$. For larger values
of j, the avalanches rapidly become much larger as is seen in
Fig. \ref{fig:comp_m_vs_h}, and for smaller values, avalanches all
become small.  Fig. \ref{fig:hyst_damping}(b) shows a plot of the same
quantity with relaxational dynamics near criticality. The avalanches
take place over a much smaller range in applied field.

\begin{figure}[htp]
\begin{center}
\includegraphics[width=\hsize]{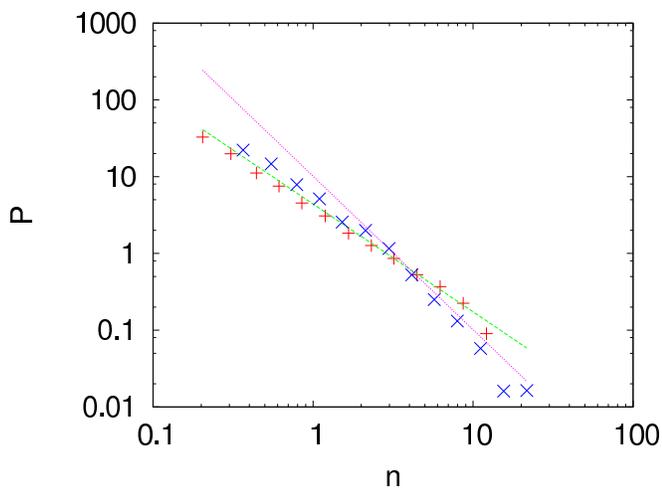}
\end{center}
\caption{
   The avalanche size distribution, measured of the entire
   hysteresis loop for $\alpha  = 0.99$ ($+$ symbols) and $\alpha = 0.9$ ($\times$ symbols).
   The x-axis is the number of avalanches normalized
   by it's mean size. The y-axis is the normalized distribution
   of sizes. The less negative sloped straight line is a fit of the $\alpha = 0.99$ curve
   and has a slope of $-1.4$. The more strongly sloped one has a slope of $-2$.
}
\label{fig:aval_dist}
\end{figure}

To quantify this difference, we studied the avalanche size distribution
exponent that is obtained by calculating the distribution of
avalanche sizes over the entire hysteresis loop. This was studied
by averaging avalanches of many runs, ($200$ for $\alpha = 0.99$)
for $32^3$ systems and for different values of parameters. We show a
comparison of the avalanche size distribution for $\alpha = 0.99$,
shown with $+$'s and for $\alpha = 0.9$, shown with $\times$'s in
Fig. \ref{fig:aval_dist}. For $\alpha = 0.99$ the curve fits quite well
to a power law with an exponent of $-1.4 \pm .1$ as shown in the figure.
For purely relaxational dynamics, the same exponent has been carefully
measured~\cite{PerkovicDahmenSethna}  to be $2.03 \pm .03$ (which is
consistent with our results for relaxational dynamics on much smaller
systems than theirs).  With smaller damping we expect to have a crossover
length corresponding to the length scale associated with the damping
time, above which the dynamics should appear relaxational.  $\alpha = 0.9$ 
appears to show such a crossover from a slope of approximately $-1.4$ 
for small avalanches, to a higher slope for large ones. A line
with slope of $-2$ is shown for comparison and appears to be consistent
with this interpretation.

\section{Discussion}

This paper has introduced a new set of dynamics for Ising models
that incorporates damping in a way that has not before been achieved.
The dynamics that have been devised have a lot in common with Langevin
dynamics, except they are for discrete rather than continuous systems.
In Langevin equations, a continuous set of stochastic differential
equations are used to model a system. It differs from molecular dynamics
in that thermal noise and damping are both added so that the system
obeys the correct equilibrium statistics.  In the case studied here,
we start by considering microcanonical dynamics~\cite{Creutz,Conti}
which introduces auxiliary degrees of freedom. We then add damping and
thermal noise. Whereas the thermal noise is typically Gaussian in the
case of the Langevin equation, here it must be taken to be of a special
exponential form, Eq.  \ref{eq:p(n)}, in order for it to satisfy the
correct equilibrium statistics.

The form of this noise, although quite unusual, can be understood,
to some extent qualitatively. For large damping, or small $\alpha$,
the strength of the $\delta$ function becomes small, and the effect
is dominated by the second term which is $\propto \exp(-\beta n)$ (for
positive $n$). Although this is non-Gaussian, $n$ can be thought of as
a random amount of positive energy. In the Langevin equation, noise is
often added to a velocity degree of freedom. In terms of a velocity, the
exponential form that we have obtained would correspond to a Gaussian
if this was expressed in terms of a velocity instead. In the limit of
small damping, where $\alpha$ is close to $1$, the effect of the noise
becomes small because the first term, which is to add no noise, will
dominate the distribution. This is in accord with what happens in the
Langevin equation where if dissipation is small, little thermal noise
is needed to keep the system at a given temperature.

The fact that it is possible to model damped systems in this discrete
manner should have many useful applications, and is easily extended
to other kinds of systems, aside from Ising models, especially in
applications where computational efficiency is an important criterion.

The case of avalanches in magnetic systems is an interesting
nonequilibrium use of these dynamics.  Although one might expect that
in most situations, for large enough distance and time scales, finite
damping will be unimportant, physics at smaller scales is still of
great interest and effects at those scales can propagate to larger scales. 
Because damping in real materials can be quite small, their
effects are readily observable experimentally. This work is expected to
be important at intermediate scales. We have investigated the phenomenon
seen in this model with varying degrees of damping and found that it
makes a qualitative difference to many of the features seen on small
and intermediate scales. This work is by no means exhaustive and there
are many other effects that can be investigated by straightforward
extensions. The effect of dipolar interactions in conjunction with
damping could also be explored. We have chosen to update the spin and
auxiliary variables at equal frequencies. Varying this should lead to a
different value for the heat diffusion coefficient which should change
the quantitative values for length and time scales.

The phenomena we have found was in qualitative agreement with earlier
work using the Landau Lifshitz Gibbs equations~\cite{DeutschBerger1}. As
avalanches progress, the effective temperature, which we have seen
can be quantified by $\langle e_i\rangle$ at site $i$, will increase
as energy is released.  This energy then diffuses to the surrounding
regions, giving those spins the opportunity to lower their energy by
thermal activation. This allows avalanches to more easily progress
when the damping is small in contrast to relaxational dynamics, which
has effectively infinite damping, $\alpha = 0$.  This can lead to some
substantial differences in avalanche morphology, particularly as for small
damping, highly disordered systems can avalanche.  At low fields this
leads to a single avalanche being composed of many disconnected pieces.
Experiments have been devised~\cite{BergerInomataEtAl} that are close
experimental realization of the two dimensional random field Ising model,
and it would be interesting to determine if systems such as this one,
or similar to it, show avalanches with this morphology.

\end{document}